\documentstyle[12pt]{article}

\def\beq{\begin{equation}}

\def\eeq{\end{equation}}

\def\bea{\begin{eqnarray}}

\def\eea{\end{eqnarray}}

\def\bq{\begin{quote}}

\def\eq{\end{quote}}

\def\nnb{\nonumber}

\def\ga{\left(}

\def\dr{\right)}

\def\rar{\rightarrow}

\def\nnb{\nonumber}

\def\la{\langle}

\def\ra{\rangle}

\def\nin{\noindent}
\def\ba{\begin{array}}
\def\ea{\end{array}}

\begin{document}

\topmargin -2.5cm

\oddsidemargin -.5cm

\evensidemargin -1.0cm

\pagestyle{empty}

\begin{flushright}
PM 95/51\\
\end{flushright}

\vspace*{0.5cm}

\begin{center}
\section*{Heavy quarkonia mass-splittings in QCD:\\
test of the $1/m$-expansion and
estimates of $\la\alpha_s G^2\ra$ and $\alpha_s$} 

\vspace*{0.5cm}
{\bf S. Narison}
\footnote{Supported in part
within the Austrian-French bilateral exchange cooperation.}
 \\
Laboratoire de Physique Math\'ematique,\\
Universit\'e de Montpellier II, Place Eug\`ene Bataillon,\\ 
34095 - Montpellier Cedex 05, France\footnote{Permanent address.}.
\\
and\\
Institut f\"ur Theoretische Physik,\\
Boltzmanngasse 5, A-1090 Wien, Austria.
\vspace*{1.5cm}

{\bf Abstract} \\ \end{center}

\nin
 {\it New double ratios} of exponential moments of different
two-point functions, 
which are less sensitive
to the heavy quark mass and to the continuum effects than
the {\it commonly used} ratio of moments, are presented 
for a
more refined analysis of the mass-splittings
between the different
heavy quarkonia states. We
 show that at the $c$ and $b$ quark mass
scales the $1/m$-expansion does not converge for these quarkonia
channels, while a connection of our mass and width formulae, 
with the potential model ones is done. Using the present
value of the QCD coupling $\alpha_s$, we deduce the value: 
$\la\alpha_s G^2\ra = (7.5\pm 2.5)\times 10^{-2}$ GeV$^4$ 
of
the gluon condensate from $M_\psi
-M_{\eta_c}$ and $M_{\chi_b}-M_\Upsilon$, which we compare
with the ones from different fits of the
heavy and light quark channels. 
We also find that 
 $M_{\chi_c(P^1_1)}-M_{\chi_c(P^3_1)}$ is gouverned by the
radiative corrections and gives
$\alpha_s$(1.3 GeV)=0.64$^{+0.36}_{-0.18}$ for 4 flavours, implying
$\alpha_s(M_Z)=0.127\pm 0.011$.
Our predictions for the splittings 
of different heavy quarkonia states 
are summarized in Table 2, where, in particular, we
find $M_{\Upsilon}-M_{\eta_b}\approx 63^{-29}_{+51}$ MeV implying the
possible observation of the $\eta_b$ in the $\Upsilon$-radiative decay.

\vspace{5cm}
\begin{flushleft}
PM 95/51\\
December 95 \\
\end{flushleft}

\vfill\eject

\setcounter{page}{1}

 \pagestyle{plain}

\subsection*{The double ratio of moments} \par
\nin
QCD spectral sum rule (QSSR) \`a la SVZ \cite{SVZ} 
(for a recent
review, see e.g. \cite{SNB}) has shown since
15 years, its impressive ability
for describing the complex phenomena of hadronic
physics with the few universal ``fundamental" parameters
 of the QCD
Lagrangian
(QCD coupling $\alpha_s$, quark masses
and  vacuum condensates built from the quarks
and/or gluon fields), without waiting for a
complete understanding of the confinement problem. 
In the example of the two-point correlator:
\beq
\Pi_Q(q^2) \equiv i \int d^4x ~e^{iqx} \
\la 0\vert {\cal T}
J_Q(x)
\ga J_Q(o)\dr ^\dagger \vert 0 \ra ,
\eeq
 associated to the generic hadronic current:
$J_Q(x) \equiv \bar Q \Gamma Q (x)$ of the heavy $Q$-quark 
($\Gamma$ is
a Dirac matrix which specifies the hadron quantum numbers),
the SVZ-expansion reads:
\bea
\Pi_Q (q^2)
&\simeq& \sum_{D=0,2,...}\sum_{dim O=D}
\frac{ C^{(J)}(q^2,M^2_Q,\mu)\la O(\mu)\ra}
{\ga M^2_Q-q^2 \dr^{D/2}}, 
\eea
where $\mu$ is an arbitrary scale that separates the long- 
and
short-distance dynamics; $C^{(J)}$ are the Wilson 
coefficients 
calculable
in perturbative QCD by means of Feynman diagrams 
techniques;
$\la O \ra$ are
the non-perturbative condensates of dimension $D$ built
 from the quarks or/and gluon
fields ($D=0$
corresponds to the case of the na\"\i ve perturbative 
contribution). Owing to gauge invariance, the lowest
dimension condensates that can be formed are the $D=4$
light quark $m_q \la\bar \psi \psi \ra$ and gluon $\la
\alpha_s 
G^2 \ra$
ones, where the former is fixed by the pion PCAC relation, 
whilst 
the latter
is known to be $(0.07\pm 0.01)$ GeV$^4$ from more recent 
analysis
of the light \cite{SNL} quark systems \cite{SNB}. 
The
validity of the SVZ-expansion has been understood 
formally, using
 renormalon
techniques (absorption of the IR renormalon ambiguity into 
the 
definitions
of the condensates and absence of
some extra $1/q^2$-terms not included 
in the OPE)
\cite{MUELLER,BENEKE} and/or by building  
renormalization-invariant
combinations of the condensates (Appendix of \cite{PICH} 
and 
references
therein). The SVZ expansion is phenomenologically 
confirmed from 
the unexpected
accurate determination of the QCD coupling $\alpha_s$ and 
from
a measurement of the condensates from semi-inclusive
$\tau$-decays \cite{PICH}--\cite{DUFLOT}.\\
The previous QCD information is transmitted to the data 
through 
the spectral function Im$\Pi_Q(t)$
via the K\"allen--Lehmann dispersion relation ($global~
duality)$
 obeyed by the hadronic correlators,
which can be improved from the uses of different versions
of the sum rules \cite{SVZ}, \cite{RRY}-\cite{BELL}. In this paper,
we shall use the simple duality ansatz parametrization:
$
``{one~narrow~resonance}"+
 ``{QCD~ continuum}"$, from a threshold $t_c$, which
gives a good description of the spectral integral in the
sum rule analysis, as has been
tested successfully in the light-quark channel 
from the 
$e^+e^- \rar$
$I=1$ hadron data and in the heavy-quark ones from the
$e^+e^- \rar \psi$ or $\Upsilon$ data. 
We shall work with the relativistic
version of the Laplace or exponential sum rules  where the QCD expression
konown to order
$\alpha_s$ is given in terms of the pole mass
$m(p^2=m^2)$ \cite{BELL},\cite{BERT}--\cite{PARK} 
\footnote{For consistency, we shall work with 
the
too-loop order $\alpha_s$ expression
of the pole mass \cite{SNM}.}:
\bea
{\cal L}(\sigma,m^2)
&\equiv& \int_{4m^2}^{\infty} {dt}~\mbox{exp}(-t\sigma)
~\frac{1}{\pi}~\mbox{Im} \Pi_Q(t)
=4m^2A_H
(\omega)\Bigg{[} 1+\alpha_s 
a_H(\omega) + \frac{\pi}{36}\frac{\la \alpha_s G^2\ra}
{m^4}b_H
(\omega) \Bigg{]},\nnb\\
{\cal R}_H(\sigma)&\equiv& -\frac{d}{d\sigma}
\log{{\cal L}_H(\sigma,m^2)}
=4m^2F_H(\omega)\Bigg{[} 1+\alpha_s 
P_H(\omega) + \frac{\pi}{36}\frac{\la \alpha_s G^2\ra}
{m^4}Q_H
(\omega) \Bigg{]},
\eea
where:
\beq
\omega \equiv 1/x = 4m^2\sigma
\eeq
is a dimensionless variable, while $\sigma \equiv \tau$ 
(notation used in the literature) is the
exponential Laplace sum rule variable; $F_H,~ P_H$ and 
$Q_H$ 
are complete QCD Whittaker functions compiled in 
Ref. \cite{BERT}--\cite{PARK}; $H$ specifies the 
hadronic channel
studied.
In principle, the pair $(\sigma,t_c)$ is free 
external
parameters in the analysis, so that the optimal result 
should be
insensitive to their variations. {\it Stability criteria}, 
which 
are equivalent
to the variational method, state that the best results 
should
be obtained at the minimas or at the inflexion points in 
$n$ or 
$\sigma$,
while stability in $t_c$ is useful to control the 
sensitivity of 
the
result in the changes of $t_c$-values. These stability 
criteria are satisfied in the heavy quark channels studied
here, as the continuum effect is negligible
and does not exceed 1\% of the ground
state contribution \cite{SNB,BERT}, such that 
at the minimum in $\sigma$, one expects
 to a good approximation:
\beq
\mbox{min}_{\sigma}{\cal R}(\sigma)\simeq M^2_H.
\eeq 
Moreover,
one can  $a ~posteriori$ check that, at the stability
point, where we have an equilibrium between the continuum 
and the
non-perturbative
contributions, which are both small,
the OPE is still convergent such that the SVZ-expansion 
makes sense. 
The previous approximation can be improved
by working with the double ratio of moments 
\footnote{This method has also been used 
in \cite{SNBOT} for studying the mass
splittings of the heavy-light quark systems.} 
:
\beq
{\cal R}_{HH'}(x)\equiv \frac{{\cal R}_H}{{\cal R}_{H'}}
\simeq \frac{M^2_H}{M^2_{H'}}
=\Delta_0^{HH'}\Bigg{[} 1+\alpha_s \Delta_{\alpha_s}^{HH'}
 + \frac{4\pi}{9}{\la \alpha_s G^2\ra}{\sigma^2x^2}
\Delta_G^{HH'} \Bigg{]},
\eeq
provided that each ratio of moments stabilizes at about 
the
same value of $\sigma$, as in this case, there is a
cancellation of the different leading terms such as the 
heavy quark mass
(and its ambiguous definition used in some previous literatures), 
the negligible continuum 
effect (which is already small in the ratio of
moments), and each leading QCD corrections. We shall 
limit 
ourselves here to the $\alpha_s$-correction for the 
perturbative contribution and to the leading order one 
in
$\alpha_s$ for the gluon condensate effects. To the order we are
working, the gluon condensate is well-defined as the ambiguity
only comes from higher order terms in $\alpha_s$, which have, however,
a smaller numerical effect than the one from the error of the
phenomenological estimate of the condensate.
\subsection*{Test of the $1/m$-expansion}

\nin
For this purpose, we use the complete {\it horrible}
results expressed in terms of the
pole mass to order $\alpha_s$ given by \cite{BERT} and checked by 
various authors
\cite{SNB}, which we expand with the help of the Mathematica 
program. We obtain for different channels the expressions given 
in Table 1. By comparing the complete and truncated series in $1/m$,
one can notice that, at the $c$ and $b$ mass scales, the convergence of 
the $1/m$-expansion is quite bad due to the increases of the numerical 
coefficients with the power of $1/m$ and to the alternate signs 
of the $1/m$ series. 
\subsection*{Balmer-mass formula from the ratio of moments}

\nin
The Balmer formula derived from a
non-relativistic approach ($m\rar \infty$)
of the Schr\"odinger levels 
reads \cite{LEUT} (see also \cite{DOSCH}--\cite{YND}).
for the $S^3_1$ vector meson:
\beq
M_{S^3_1}\simeq 2m
\Bigg{[}1-\frac{2}{9}\alpha_s^2+
0.23\frac{\pi}
{(m\alpha_s)^4}\la\alpha_s G^2\ra\Bigg{]}.
\eeq
\begin{table*}
\setlength{\tabcolsep}{1.5pc}
\caption{Expanded expressions of different QCD corrections
in the case of the pole mass $m(p^2=m^2)$ known to order $\alpha_s$.}
\begin{center}
\begin{tabular}{c|c}
&\\
\hline
\hline
&\\
&Vector $S^3_1$\\
&\\
$\pi A_V$&$\frac{3}{16\sqrt{\pi}}x^{3/2}\ga 1-\frac{3}{4}x+
\frac{45}{32}x^2-\frac{525}{128}x^3+...\dr$\\
$a_V$&$\frac{4}{3\sqrt{\pi}}\ga \frac{\pi}{\sqrt{x}}+0.040+
1.952\sqrt{x}-1.539x-..\dr.$\\
$b_V$&$-\frac{1}{2x^3}+\frac{3}{2x^2}+\frac{27}{8x}-
\frac{21}{8}+...
$\\
$F_V$&$1+\frac{3}{2}x-\frac{5}{4}x^2+5x^3-...$\\
$P_V$&$-\frac{2}{3}\sqrt{\pi x}+2.704x^{3/2}-10.093x^{5/2}+
52.93x^{7/2}-
...$\\
$Q_V$&$\frac{3}{2x^2}-\frac{1}{4x}+\frac{13}{8}-\frac{41}
{4}x
+...$\\
&\\
\hline
&\\
&S-waves splitting\\
&\\
$\Delta_0^{VP}$&$ 1-\frac{x^2}{2}+\frac{7}{2}x^3-...$\\
$\Delta_\alpha^{VP}$&$ \frac{\sqrt{\pi}}{9}x^{3/2}+1.539
x^2
-3.0258x^{5/2}
-7.719x^3+26.307x^{7/2}+...$\\
$\Delta_G^{VP}$&$ \frac{5}{x}\ga
1-\frac{4}{5}x+\frac{11}{10}{x^2}+\frac{17}{10}x^3-...
\dr$\\
&\\
\hline
&\\
&P-waves splittings\\
&\\
$\Delta^{01}_0$&1\\
$\Delta^{01}_\alpha$&$-3.18x^2(1-10.17x+102.1x^2+...)$\\
$\Delta^{01}_G$&$ -\frac{2}{x}+\frac{5}{2}-\frac{55}{4}x
+...
$\\
$\Delta^{13}_\alpha$&$1.06x^2(1-9.5x+81.1x^2-...)$\\
$\Delta^{AT}_0$&$1+x^2-\frac{23}{2}x^3+...$\\
$\Delta^{AT}_\alpha$&$-0.1576x^{3/2}-2.545x^2+3.95x^{5/2}
-...
$\\
$\Delta_G^{AT}$&$ -\frac{6}{x}+\frac{31}{4}-\frac{89}{8}x
-
\frac{1715}{8}x^2+...$\\
&\\
\hline
&\\
&P- versus S-waves splittings\\
&\\
$\Delta^{VS}_0$&$1-x+5x^2-30x^3+...$\\
$\Delta^{VS}_\alpha$&$-\frac{2}{9}\sqrt{\pi x}-0.336x^{3/2}
+4.244x^2+
7.458x^{5/2}-42.017x^3-...$\\
$\Delta^{VS}_G$&$-\frac{3}{x^2}-\frac{2}{x}-\frac{41}{4}+
\frac{389}{4}x-...$\\
$\Delta^{VA}_0$&$\Delta^{VS}_0$\\
$\Delta^{VA}_\alpha$&$-\frac{2}{9}\sqrt{\pi x}-0.336x^{3/2}
+1.06x^2+
7.458x^{5/2}-9.655x^3...$\\
$\Delta^{VA}_G$&$-\frac{3}{x^2}-\frac{4}{x}-\frac{31}{4}+
\frac{167}{2}x-...$\\
$\Delta^{VT}_0$&$1-x+6x^2-\frac{85}{2}x^3+...$\\
$\Delta^{VT}_\alpha$&$-\frac{2}{9}\sqrt{\pi x}-0.493x^{3/2}
-1.484x^2
+11.409x^{5/2}+18.248x^3-...
$\\
$\Delta^{VT}_G$&$-\frac{3}{x^2}-\frac{10}{x}+\frac{579}{8}x
-\frac{16719}{16}x^2+...$\\
&\\
\hline
\hline
\end{tabular}
\end{center}
\end{table*}

\nin
It is instructive to compare this result with the mass
formula obtained from the ratio of moments 
within the $1/m$-expansion. Using the different QCD
corrections in Table 1, one obtains the mass formula
at the minimum in $\sigma$ of ${\cal R}$:
\beq
M_{S^3_1}\simeq \sqrt{{\cal R}(\sigma_{min})}\simeq 
2m(1+\frac{3}{16m^2\sigma})
\Bigg{[}1-\frac{\sqrt{\pi}}{6m}\frac{\alpha_s(\sigma)}
{\sqrt{\sigma}}
+\frac{\pi}{12}\sigma^2\la\alpha_s G^2\ra\Bigg{]}.
\eeq
In the case of the $b$-quark, where we expect the 
{\it static
approximation} more reliable,the minimum of this quantity 
is obtained to leading order at:
\beq
\sqrt{\sigma_{coul}}\simeq \frac{9}{4m\alpha_s\sqrt{\pi}}\simeq
0.85 ~\mbox{GeV}^{-1}~,
\eeq
where we have used for 5 flavours \footnote{In the approximation 
we are working, the effect of the number of flavours enters only
through the $\beta$-function and therefore is not significant.}:
$\alpha_s(\sigma)\simeq 0.32\pm 0.06$
 after evolving the  value $\alpha_s(M_Z)=0.118\pm 
0.006$ from LEP \cite{BETHKE} and  $\tau$-decay data
\cite{PICH}--\cite{DUFLOT}.  
The inclusion of the gluon condensate correction shifts 
the value of $\sigma_{min}$ to:
\beq
\sqrt{\sigma_{min}}\simeq 0.74\sqrt{\sigma_{coul}}.
\eeq
These previous values of $\sigma$ confirm the more involved
numerical analysis in \cite{BERT}
and indicates the relevance of the gluon condensate
in the analysis of the spectrum.
By introducing the previous leading order
expression of $\sigma_{coul}$
into the sum rule, one obtains:
\beq
M_{\Upsilon}\simeq 2m_b\ga 1+\frac{\pi}{27}\alpha_s^2\dr
\Bigg{[}1-\frac{2}{9}\ga\frac{\pi}{3}\dr\alpha_s^2+
\ga \frac{27}{128}\dr\ga\frac{3}{\pi}\dr^2\frac{\pi}
{(m_b\alpha_s)^4}\la\alpha_s G^2\ra\Bigg{]}.
\eeq
where one can deduce by identification in the 
{\it static limit}
($m_b\rar \infty$) that the 
 Coulombic effect is exactly the same in the two 
approaches. The apparent factor $\pi/3$
is due to the fact that we use here the approximate Schwinger 
interpolating formula for the two-point correlator . 
The gluon 
condensate coefficient is also about the 
same in the two approaches. This agreement indicates the
consistency of the potential model and sum rule approach 
in the static limit, though a new extra $\alpha_s^2$ correction 
due to the 
$v^2$ (finite mass) terms in the
free part appears here
 (for some derivations of the relativistic correction
in the potential approach see \cite{BYERS}, \cite{ZAL2}), 
and tends to reduce 
the coulombic interactions. On the other hand, at the 
$b$-quark
 mass scale, the dominance of the gluon condensate contribution 
indicates that the $b$-quark is not enough heavy
for this system to be coulombic rendering
the non-relativistic potential approach to be a
crude approximation at this scale. 
\subsection*{$S^3_1-S^1_0$ hyperfine and $P-S$-wave 
splittings} 

\nin
In the non-relativistic approach \`a la \cite{YND}, 
the hyperfine 
and $S-P$ wave splittings
are given to leading order by:
\bea
M(S^3_1)-M(S^1_0)&\simeq& 2m\frac{(C_F\alpha_s)^4}{6}
\Big{[}1+3.255\frac{\pi}
{2m^4\alpha_s^6}\la\alpha_s G^2\ra\Bigg{]}\nnb\\
M(P^3_1)-M(S^1_3)&\simeq& 2m\Bigg{[}\frac{3(C_F
\alpha_s)^2}{32}
+\frac{2^5\pi}
{(C_Fm\alpha_s)^4}\la\alpha_s G^2\ra\Bigg{]},
\eea
where $C_F=4/3$.
Using the double ratio of moments and the QCD corrections
given in Table 1, one obtains at $\sigma_{coul}$:
\bea
\frac{M(S^3_1)-M(S^1_0)}{M(S^3_1)}&\approx& -4\pi^2
\ga\frac{\alpha_s}{9}\dr^4+\frac{8}{9}
\ga\frac{\sqrt{\pi}\alpha_s}{9}\dr^3\alpha_s+
\frac{45}{32m^4\alpha_s^2}\la\alpha_s G^2\ra +...\nnb\\
\frac{M(P^3_1)-M(S^3_1)}{M(S^3_1)}&\approx&
\frac{ 4\pi}{81}\alpha_s^2+\frac{2\pi}{81}
\alpha_s^2+
\frac{27}{8m^4\alpha_s^2}\la\alpha_s G^2\ra +...,
\eea
where the corrections are, respectively, relativistic,
Coulombic and non-perturbative.
By comparing the sum rules in Eqs. (8) and (13), one can
realize that the leading $x$ or $1/\sigma$-terms cancel 
in the hyperfine
splitting, while the $x$-expansion is slowly convergent
for the $\alpha_s$-term at the $b$-mass.
Comparing now this result with the one from the 
non-relativistic approach, it is interesting to notice 
that
both approaches lead to the same $\alpha_s$-behaviour 
of the
Coulombic and gluon condensate contributions. A one 
to one correspondence of each QCD 
corrections is not very conclusive, and needs
an evaluation of the correlator at the next-next-to-leading 
order for a better control of the $\alpha^2_s{x}$ terms.  
However, as the Coulombic potential is a 
fundamental aspect of QCD, 
 we shall, however, expect that, after the resummation 
of the higher order terms in $\alpha_s$,
the coefficient of the $\alpha_s^4$-term in
the hyperfine splitting will be the same in the two
alternative approaches.
In the case of the $S-P$ wave splitting, the sum of
the $\alpha_s^2$ corrections agrees from the
two methods, though one can also notice that the
relativistic correction is larger than the Coulombic
one. 
The discrepancy for the coefficients of the gluon 
condensate 
in the two approaches is more subtle and may reflect
the difficulty of Bell-Bertlmann \cite{BERTM} to find a 
bridge between the field theory \`a la SVZ
(flavour-dependent confining potential) 
and the potential models (flavour-independence). 
Resolving the diffferent
puzzles encountered during this comparison is outside the scope 
of the present paper.
\subsection*{Leptonic width and quarkonia wave 
function }
Using the sum rule ${\cal L}_H$ and saturating it by 
the vector $S^3_1$ state, we obtain, to a good approximation,
 the sum rule:
\beq
M_V\Gamma_{V\rar e^+e^-}\simeq {(\alpha e_Q)^2}
\frac{e^{2\delta m M_V \sigma}}{72\sqrt{\pi}}
\frac{\sigma^{-3/2}}{m}
\Bigg{[} 1+\frac{8}{3}\sqrt{\pi\sigma}m\alpha_s
-\frac{4\pi}{9}\la \alpha_s G^2\ra m\sigma^{5/2}
\Bigg{]}~, 
\eeq
where $e_Q$ is the quark charge in  units of e; $\delta 
m \equiv
M_V-2m$ is the meson-quark mass gap. In 
 the case of the $b$-quark, we use \cite{SNM} $\delta m \simeq 
0.26 $ GeV,
and the value of $\sigma_{min}$ given in Eq. (10). Then:
\beq
\Gamma_{\Upsilon(S^3_1)\rar e^+e^-}\simeq 1.2~\mbox{keV}~,
\eeq
in agreement with the data 1.3 keV. However, one should remark 
from Eq. (14), that the $\alpha_s$ correction 
is huge and needs an evaluation of the higher
order terms (the gluon condensate effect is negligible),
while the exponential factor effect is large, such that 
one can
{\it reciprocally} use the data on the width to fix either 
$\alpha_s$ or/and the quark mass. Larger value of the 
heavy quark
mass at the two-loop level
(see e.g. \cite{VOL1})
corresponding to a negative value of $\delta m$, would imply a 
smaller
value of the leptonic width in disagreement with the data.

\nin
In the non-relativistic approach,
one can express the quarkonia
leptonic width in terms of its wave function $\Psi(0)_Q$:
\beq
\Gamma_{V\rar e^+e^-}=\frac{16\pi\alpha^2}
{M^2_V}e^2_Q |\Psi(0)|^2_Q\ga 1-4C_F\frac{\alpha_s}
{\pi}\dr~,
\eeq
where (see e.g. \cite{YND}):
\beq
16\pi|\Psi(0)|^2_Q \ga 1-4C_F\frac{\alpha_s}{\pi}\dr
\simeq 2(mC_F\alpha_s)^3~\approx 15~\mbox{GeV}^3.
\eeq 
In our approah, one can deduce:
\bea
16\pi|\Psi(0)|^2_Q \ga 1-4C_F\frac{\alpha_s}{\pi}\dr
&\simeq& \frac{1}{72\sqrt{\pi}}
e^{2\delta m M_V \sigma}
{\sigma^{-3/2}}\frac{M_V}{m}
\Bigg{[} 1+\frac{8}{3}\sqrt{\pi\sigma}m\alpha_s
-\frac{4\pi}{9}\la \alpha_s G^2\ra m\sigma^{5/2}
\Bigg{]}\nnb\\
&\simeq& 18.3~\mbox{GeV}^3~. 
\eea
Using the expression of $\sigma_{coul}$, one can find 
that,
to leading order, the two approaches give 
a similar behaviour  for $\Psi(0)_Q$
in $\alpha_s$ and in $m$
 and about the same value of this quantity, 
though, one should notice that
in the present approach, the
QCD coupling $\alpha_s$ is evaluated at the scale $\sigma$
as dictated by the renormalization group equation obeyed
by the Laplace sum rule \cite{SNR} but $not$ at the resonance mass!
\subsection*{ Gluon condensate
from $M_{\psi(S^3_1)}-M_{\eta_c(S^1_0)}$}

\nin
The value of $\sigma$, at which, the $S$-wave charmonium 
ratio of sum rules stabilize is: 
\cite{BERT}:
\beq
\sigma \simeq (0.9\pm 0.1) ~\mbox{GeV}^{-2}.
\eeq
Using the range of the charm quark  
pole mass to order $\alpha_s$
accuracy \cite{SNM} \footnote{For a recent 
review on 
the heavy quark masses, see e.g. \cite{ROD,PDG}.}:
$
m_c\simeq 1.2-1.5~\mbox{GeV}
$
one can deduce the conservative value of $x$:
\beq
\omega \equiv 1/x \simeq 6.6\pm 1.8~.
\eeq 
The ratio of the mass squared of the vector 
$V(S^3_1)$ and the pseudoscalar $P(S^1_0)$ 
is controlled by the double ratio of moments 
given generically in Eq. (6), where the exact 
expressions of the corrections read:
\bea
\Delta_0^{VP} \simeq 0.995^{+ 0.001}_{-0.004}~~~
\Delta_\alpha^{VP} \simeq
~0.0233^{- 0.009}_{+0.011}~~~
\Delta_G^{VP}\simeq 29.77^{+ 8.86}_{-10.23},
\eea
where each terms lead to be about 0.5, 2 and 7 $\%$ of 
the leading order one. One can understand from the approximate 
expressions in Table 1 that the leading $x$-corrections
appearing in the ratio of moments cancel in
the double ratio, while the remaining
corrections are relatively small. 
However, the $x$-expansion is not convergent for
the $\alpha_s$-term at the charm mass, which invalidates
the use of the $1/m$-expansion  done
in \cite{DOMI} in this channel.
Using for 4 flavours \cite{SNM}: $
\alpha_s(\sigma)\simeq 0.48^{+ 0.17}_{-0.10}$,
and the experimental data \cite{PDG}:
${\cal R}^{exp}_{VP}= 1.082,$
one can deduce the value of the gluon condensate:
\beq
\la \alpha_s G^2\ra \simeq (0.10\pm 0.04)~\mbox{GeV}^4.
\eeq
We have estimated the error due to higher order effects 
by replacing the coefficient of $\alpha_s$ with the one
obtained from the effective Coulombic potential, which
tends to reduce the estimate to 0.07 GeV$^4$.
We have tested the convergence of the QCD series in 
$\sigma$,
by using the numerical estimate of the dimension-six
gluon condensate $g\la f_{abc}G^aG^bG^c\ra$ contributions
given in \cite{PARK}. This effect is about 0.1$\%$
of the zeroth order term and does not influence
the previous estimate in Eq. (22), which also indicates 
the
good convergence of the ratio of exponential moments 
already
at the charm mass scale in contrast with the $q^2=0$ 
moments
studied in Ref. \cite{SVZ,RADYU}. We also expect that in 
the double ratio of moments used here, the radiative 
corrections to the gluon condensate effects 
(their expression for the
two-point correlator is however available in the
literature \cite{BROAD})
are much smaller than in the
individual moments , such that they will give a 
negligible 
effect in the estimate of the gluon condensate.  
This value obtained at the same level of
$\alpha_s$-accuracy as previous sum rule results, 
confirm the
ones of Bell-Bertlmann
\cite{BELL,BERT,PARK,BROAD,SNB} from the ratio of 
exponential moments and from FESR-like sum rule
for quarkonia \cite{MILLER,SNB} claiming that
the SVZ value \cite{SVZ} has been underestimated by
about a factor 2 (see also \cite{RADYU,ZAL}). 
Our value is also
in agreement with the more recent estimate 
$(0.07\pm 0.01)~\mbox{GeV}^4$ from the
$\tau$-like sum rules \cite{SNL}, and 
FESR \cite{FESR2} in $e^+e^-\rar I=1$ hadrons. A more 
complete comparison of different determinations is done 
in Table 2.

\begin{table*}
\setlength{\tabcolsep}{1.5pc}
\caption{Predictions for the gluon condensate, for the 
different
mass-splittings (in units of MeV) and for the leptonic 
widths
(in units of keV). 
We use $\alpha_s(M_Z)=0.118\pm 0.006$ from LEP and 
$\tau$-decay.} 
\begin{center}
\begin{tabular}{c|c|c|c}
Observables&Input&Predictions&Data / comments\\
\hline
\hline
&&&\\
$\la \alpha_s G^2\ra$[GeV]$^4\times 10^{2}$
&$M_\psi-M_{\eta_c}=108
$&$10\pm 4$& This work \\
&$M^{c.o.m}_{\chi_b}-M_{\Upsilon}=440$&$6.5\pm 2.5$&-- \\
&{ Average}&${ 7.5\pm 2.5}$ & 
Mass splittings\\
&&&\\
& Charmonium masses&
$\approx 4$ &SVZ-value \cite{SVZ}\\
&&&$q^2=0$-mom.\\
&--&$5.3\pm 1.2$ &$q^2$-mom. \cite{RRY}\\
&--&$10\pm 2$& exp. mom.\cite{BELL,BERT}\\
&--&$9.2\pm 3.4$& mom.\cite{MILLER}\\
&&&\\
&$e^+e^-\rar I=1$ hadrons&$4\pm 1$&ratio of mom.
\cite{LNT}\\
&--&$13^{+5}_{-7}$&FESR \cite{FESR2}\\
&--&$7\pm 1$&
$\tau$-like decay  \cite{SNL}\\
&$\tau$-decay (axial)&$6.9\pm 2.6$& \cite{SOLA}\\
&&&\\
&$\tau$-decay&&data\\
&ALEPH&&$7.5\pm 3.1$ \cite{DUFLOT} \\
&CLEO&&$2.0\pm 3.8$ \cite{DUFLOT}\\
\hline
&&&\\
$\alpha_s(1.3~\mbox{GeV})$&$M_{\chi_c(P^1_1)}-M_{\chi_c(P^3_1)}$&
$0.45^{+0.18}_{-0.29}$&$\alpha_s(M_Z)\simeq 0.124\pm 0.012$ \\
\hline
&&&\\
$M_{\chi_c(P^1_1)}-M_{\chi_c(P^3_1)}$&$\alpha_s$ from LEP/$\tau$-decay&
$10.1^{-4.1}_{+9.9}$
& 11.1 (c.o.m)\\
&&&\\
$M_{\chi_c(P^3_1)}-M_{\chi_c(P^3_0)}$&
$\la \alpha_s G^2\ra$~average&$89^{-16}_{+35}$& 95 \\
$M_{\chi_c(P^3_2)}-M_{\chi_c(P^3_1)}$&--&$77^{+26}_{-11}$
& 50\\
\hline
&&&\\
$M_\Upsilon-M_{\eta_b}$&$\la \alpha_s G^2\ra$~average 
&$ 13^{-7}_{+10}$& order $\alpha_s$\\
&&$63^{-29}_{+51}$&coef. $\alpha_s$: Coul. pot.\\ 
$M_{\chi_b(P^3_0)}-M_\Upsilon$&--&$ 475^{+75}_{-64}$&400\\
$M_{\chi_b(P^3_1)}-M_\Upsilon$&--&$ 485^{+25}_{-68}$&
432\\
$M_{\chi_b(P^3_2)}-M_\Upsilon$&--&$ 500\pm 71$&453\\
$M_{\chi_b(P^1_1)}-M_{\Upsilon}$&c.o.m. &
$ 492^{+56}_{-69}$& 440\\
\hline
&&&\\
$M_{\cal T}-2m_t$&$\la \alpha_s G^2\ra$~average &--906&
 two-loop pole mass\\
$M_{\cal T}-M_{\eta_t}$&--&1.8 & order $\alpha_s$\\
&&93&coeff. $\alpha_s$: Coul. pot.\\
$M_{\chi_t}-M_{\cal T}$&--&1800 &--\\
\hline
&&&\\
$\Gamma_{\Upsilon\rar e^+e^-}$&--&1.2& 1.32 \\
$\Gamma_{{\cal T}\rar e^+e^-}$&--&0.16 &--\\
\hline
\hline
\end{tabular}
\end{center}
\end{table*}
\subsection*{Charmonium $P$-wave splittings}

\nin
The analysis of the different ratios of moments for the
$P$-wave charmonium shows 
\cite{BELL}-\cite{PARK} that they are optimized for:
\beq
\sigma\simeq (0.6\pm 0.1)~\mbox{GeV}^{-2},~~~~~~~~\Longrightarrow~~~~~~~~~
\alpha_s(\sigma)\simeq 0.41^{+ 0.11}_{-0.07}~,~~~~~~~~~~~
1/x=4.5\pm 1.5.
\eeq

\nin
{\bf In the case of the Scalar $P^3_0$ - axial $P^3_1$ mass 
splitting}, the different exact QCD coefficient corrections of 
the corresponding double ratio of moments read:
\bea
\Delta^{01}_0=1,~~~~~~
\Delta^{01}_\alpha\simeq -(0.045^{- 0.014}_{+0.028}),
~~~~~~
\Delta^{01}_G\simeq -(7.75^{+ 2.84}_{-2.77}).
\eea
Using the correlated values of the different parameters, 
one
obtains the mass-splitting
$
\Delta M^{3}_{10}\equiv
M_{P^3_1}-M_{P^3_0}\simeq (60^{- 16}_{+35})~\mbox{MeV},
$
where we have used the experimental value $M_{P^3_1}=
3.51$
GeV. Adding the $\la g G^3\ra$ dimension-six
condensate effect,which
is about -1.6$\%$ of the leading term in ${\cal R}_{01}$, one
can finally deduce the prediction in Table 2, which is
in excellent agreement with the data .
One should remark that the previous predictions indicate 
that, for the method to reproduce correctly the 
mass-splittings of the $S$ and $P$-wave charmonium states, one
needs {\it both} larger values of $\alpha_s$ and $\la
\alpha_s G^2\ra$ than the ones favoured in the early days 
of the sum rules.

\nin  
{\bf In the case of the Tensor $P^3_2$-axial $P^3_1$ mass 
splitting},
the different exact QCD corrections for the
double ratio of the tensor over the axial meson
moments read:
\beq
\Delta_0^{TA}=(0.989^{+0.003}_{-0.006})~~~~
\Delta_\alpha^{TA}=(0.029^{-0.004}_{+0.013})~~~~
\Delta_G^{TA}=(22.1^{+8.5}_{-8.2}),
\eeq
from which, one can 
deduce
the prediction in Table 2, which is slightly higher than 
the data of 50 MeV. This small
discrepancy may be attributed to the unaccounted effects of the 
dimension-six condensate or/and to the (usual) increasing role
of the continuum for state with higher spins. However, 
the prediction is quite satisfactory within our approximation.

\nin
\subsection*{\bf $\alpha_s$ from the $P^1_1$ - $P^3_1$ 
axial mass splitting}
 The corresponding double ratio of moments has the nice 
feature 
to be independent of the gluon condensate to leading 
order in $\alpha_s$ and reads:
\beq
\frac{M_{P^1_1}^2}{M_{P^3_1}^2}\simeq 1+\alpha_s\Bigg{[}
\Delta_\alpha^{13}(\mbox{exact})=0.014^{-0.004}_{+0.008}\Bigg{]}~.
\eeq
The recent experimental value of the $P^1_1$ state denoted by
$h_c(1P)$ in the PDG compilation \cite{PDG} coincides with
the one of the center of mass energy: 
\beq
M_{P^1_1}\simeq M^{c.o.m}_P=\frac{1}{9}\Bigg{[} 
5M_{P^3_2}+3M_{P^3_1}+M_{P^3_0}\Bigg{]}\simeq 11.1~\mbox{MeV}
\eeq
as expected from the short range nature of the spin-spin force.
We use the experimental value of the $h_c(1P)$ mass of 3526.1 MeV, 
and a na\"{\i}ve exponential
resummation of the higher order $\alpha_s$ terms. Then, we deduce:
\beq 
\alpha_s(\sigma^{-1}\simeq
1.3~\mbox{GeV})\simeq 0.64^{+0.36}_{-0.18}\pm 0.02~~~~~\Longrightarrow
~~~~~\alpha_s(M_Z)\simeq 0.127\pm 0.009 \pm 0.002                                                                                           \pm 0.032\pm 0.030,
\eeq
where
The error is much bigger than the one from LEP and $\tau$ decay data,
but its value is perfectly consistent with the later. The theoretical error
is mainly due to the uncertainty in $\Delta_{\alpha}$, while a na\"{\i}ve
resummation of the higher order $\alpha_s$ terms leads the second error. 
However, though inaccurate,
this value of $\alpha_s$ is interesting for
an alternative derivation of this fundamental quantity at lower energies, 
which can
serve for testing its $q^2$-evolution until $M_Z$.\\
{\it Reciprocally}, using the value of $\alpha_s$ from LEP and 
$\tau$-decay
data as input, one can deduce the prediction of the center of 
mass (c.o.m) of the $P^3_J$ states given in Table 2. 

\subsection*{$\Upsilon-\eta_b$ mass splitting}

\nin
For the bottomium, the analysis of the ratios of moments 
for the 
$S$ and $P$ waves shows that they are optimized at the 
same
value of $\sigma$, namely \cite{BERT}:
\beq
\sigma= (0.35\pm 0.05)~\mbox{GeV}^{-2},
\eeq
which implies for 5 flavours: $
\alpha_s(\sigma)\simeq 0.32\pm 0.06$.
Using the conservative values of the two-loop
$b$-quark pole mass:
$m_b \simeq 4.2-4.7 ~\mbox{GeV}$,
one can deduce:
\beq
1/x\simeq 28\pm 7~,
\eeq
where one might (a priori) expects a good convergence of the $1/m$ 
expansion. 

\nin
The splitting between the vector $\Upsilon(S^3_1)$ and the
pseudoscalar $\eta_b(S^1_0)$ can be done in a similar
way than the charmonium one. The double ratio of moments
reads numerically:
\beq
{\cal R}_{VP}\simeq \frac{M^2_V}{M^2_P}\simeq
(0.9995^{+0.0002}_{-0.0003})\Bigg{[}1+\alpha_s\ga 
2.4^{-0.7}_
{+1.4}\dr\times 10^{-3}+(0.03\pm 0.01)\mbox{GeV}^{-4}
\la\alpha_s G^2\ra \Bigg{]},
\eeq
where we have used the exact expressions of the QCD 
corrections. It leads to the mass splitting in Table 2.
To this order of perturbation theory, 
this result is in the range of the potential model 
estimates
\cite{MARTIN,BUCH,YND}, with the exception of the one in
\cite{LEUT} and \cite{BYERS}, where in the latter it has 
been shown that the square of the quark velocity $v^2$ 
correction
can cause a large value of about 100 MeV for the 
splitting.
One should also notice that, to this approximation, 
the gluon
condensate gives still the dominant effect at the $b$-mass 
scale (0.2$\%$ of the leading order) compared
to the one $.08\%$ from the $\alpha_s$-term. 
However, the $1/m$ series of the QCD $\alpha_s$
correction is badly convergent, showing that the static limit 
approximation is quantitavely inaccurate in the $b$-channel. Therefore, 
one expects that the corresponding prediction of $(13^{-7}_{+10})$ MeV 
is a very crude estimate. In order to 
control the effect of the unknown higher order terms,
it is legitimate to introduce into the sum rule,
the coefficient of the 
Coulombic effect from the QCD potential as given by the
$\alpha_s^2$-term in Eq. (12)\footnote{In this case, the 
gluon condensate contribution
is smaller than the Coulombic one as has been observed
in \cite{GIACO}.}. Therefore, we deduce the ``improved" 
final estimate in Table 2:
\beq
M_{\Upsilon}-M_{\eta_b}\approx \ga 63^{-29}_{+51}\dr~\mbox{MeV},
\eeq
implying the possible observation of the $eta_b$ from the $Upsilon$ 
radiative decay.
\subsection*{$\Upsilon-\chi_b$ mass splittings and new 
estimate
of the gluon condensate}

\nin
As the $S$ and $P$ wave ratios of moments are optimized 
at the
same value of $\sigma$, we can compare directly, with a 
good
accuracy, the different $P$ states with 
the $\Upsilon~(S^3_1)$ one. As the coefficient of the 
$\alpha_s^2$ corrections, after inserting the expression
of $\sigma_{min}$, 
are comparable with the one from the Coulombic
potential, we expect that the prediction of this 
splitting
is more accurate than in the case of the hyperfine.
The different double ratios of
moments read numerically for the values in Eqs (28)--(29):
\bea
{\cal R}_{VS}&\simeq& \frac{M^2_V}{M^2_S}\simeq
(0.9696^{+0.0054}_{-0.0083})\Bigg{[}1-\alpha_s
(0.071^{-0.006}_
{+0.011})-(0.50^{+0.18}_{-0.11})\mbox{GeV}^{-4}
\la\alpha_s G^2\ra \Bigg{]},
\nnb\\
{\cal R}_{VA}&\simeq& \frac{M^2_V}{M^2_A}\simeq
(0.9696^{+0.0054}_{-0.0083})\Bigg{[}1-\alpha_s
(0.074^{-0.007}_
{+0.012})-(0.54^{+0.18}_{-0.12} )\mbox{GeV}^{-4}
\la\alpha_s G^2\ra \Bigg{]},
\nnb\\
{\cal R}_{VT}&\simeq& \frac{M^2_V}{M^2_T}\simeq
(0.9704^{+0.0051}_{-0.0084})\Bigg{[}1-\alpha_s
(0.077^{-0.008}_
{+0.006})-(0.57^{+0.16}_{-0.13} )\mbox{GeV}^{-4}
\la\alpha_s G^2\ra \Bigg{]},
\eea
where $V,~S,~A,~T$ refer respectively to the $\Upsilon$ 
and to the different $\chi_b$ states $P^3_0,~
P^3_1,~P^3_2$. Using the value of the
gluon condensate obtained previously, 
these sum rules lead to the 
mass-splittings in Table 2, which is in good agreement 
with the
corresponding data, but 
definitely higher than the previous predictions of
\cite{VOL}, where, among
other effects, the values of $\alpha_s$ and of the gluon
condensate used there are too low.
Reciprocally, one can use the data for
re-extracting {\it independently} the value of the gluon
condensate. As usually observed in the literature, the 
prediction
is more accurate for the c.o.m., than for the individual
mass. The corresponding numerical sum rule is:
\beq
\frac{M_{\chi_b}^{c.o.m}-M_{\Upsilon}}{M_{\Upsilon}}
\simeq 
\ga 1.53^{+0.26}_{-0.42}\dr \times 10^{-2}+
\ga 1.20^{+0.1}_{-0.2}\dr \times 10^{-2}
+(0.28^{+0.08}_{-0.06} )\mbox{GeV}^{-4}
\la\alpha_s G^2\ra~, 
\eeq
which leads to:
\beq
\la \alpha_s G^2\ra \simeq (6.9\pm 2.5)\times 
10^{-2}~\mbox{GeV}^4.
\eeq
We expect that this result is more reliable than the one 
obtained
from the $M_\psi-M_{\eta_c}$ as the latter can be more
affected by the non calculated next-next-to-leading 
perturbative
radiative corrections than the former.
An average of the two
results from the $\psi-\eta_c$ and $\Upsilon-\chi_b$
mass splittings
leads to:
\beq
\la \alpha_s G^2\ra \simeq (7.5\pm 2.5)\times 
10^{-2}~\mbox{GeV}^4,
\eeq
where we have retained the most precise error. This result
can be compared with different fits
 of the heavy and 
light quark channels given in Table 2, and which range from
4 (SVZ) to 14 in units of $10^{-2}~\mbox{GeV}^4$. The 
most recent estimate from $e^+e^-\rar I=1$ hadrons data using
$\tau$-like decay is $(7\pm 1)\times 
10^{-2}~\mbox{GeV}^4$, where one
should also notice that the different post-SVZ estimates
favour higher values of the gluon condensate. More accurate
measurements of this quantity than the already available
results from $\tau$-decay data \cite{DUFLOT} are needed for
testing the previous phenomenological estimates from the
sum rules.

\subsection*{Toponium: illustration of the infinite mass limit}
Because in the alone case of the toponium, the $1/m$-expansion
is ideal, we have extended the previous analysis in this channel,
though, we are aware that this application can be purely academic 
because of
the eventual inexistence of the corresponding bound states.
We use the top mass:
$
m_t \simeq (173\pm 14)~\mbox{GeV},
$
obtained from the average of the CDF candidates 
$(174\pm 16)$ GeV and of the electroweak
data $(169\pm 26)$ GeV as compiled by PDG \cite{PDG}.
We shall work with the ratio of moments in the vector 
channel for determining the mass of the $S^3_1$ state,
and use with a good confidence the leading terms of the 
expressions
given in Table 1. Using the sum rules
in Eqs. (8) and (11) and the value of the minimum $\sigma^{-1/2}\simeq
20~\mbox{GeV}$ from Eq. (9), we deduce
the result for the meson-quark mass gap
given in Table 2.
 For the splittings, we use the sum rules in Eqs. (12) and (13), 
while, for the leptonic width, we use the sum rule in Eq. (14).
Our results are summarized in Table 2.
\subsection*{Conclusions}
We have used {\it new double ratios} of exponential sum 
rules
for directly extracting  the mass-splittings of different 
heavy quarkonia states. Therefore, we have obtained
from $M_\psi-M_{\eta_c}$ and $M_{\chi_b}-M_{\Upsilon}$
a more precise estimate of
the value of the gluon condensate given in Eq. (34). We
have also used $M_{\chi_c(P^1_1)}-M_{\chi_c(P^3_1)}$ 
for an alternative
extraction of $\alpha_s$ at low energy (see Eq. (28)), 
with a value consistent
with the one from LEP and $\tau$-decay.
Our numerical results are summarized in Table 2, 
where a comparison
with different estimates and experimental data is done. We
have not extended this analysis to the $D$-waves as these
states are described with operators of higher dimensions,
where, in general, the role of the continuum is 
relatively important and can obtruct the extraction of the 
mass-splittings.
\nin
We have also attempted to connect the sum rules and the
potential model approaches, using a $1/m$-expansion. We
found, that the Coulombic corrections, which are 
quite well
understood in QCD, agree, in general,
in the two approaches, expect in the
radiative corrections of the hyperfine splitting which 
requires
the knowledge of the next-next-to-leading 
$\alpha_s$-corrections.
Relativistic corrections due to finite value of the
quark mass have been included in our analysis.
However, the coefficients of the gluon condensate disagree 
in the two approaches, which may be related to the difficulty 
encountered 
by Bell-Bertlmann in finding a bridge between a
field theory \`a la SVZ and potential models.
\subsection*{Acknowledgements}
It is a pleasure to thank the French Foreign Ministry, the
French Ambassy in Vienna and the Austrian Ministry of 
Research for a financial support within a
bilateral scientific cooperation, the 
Institut f\"{u}r Theoretische Physik of
Vienna for a hospitality, Reinhold
Bertlmann for a partial collaboration in the early stage 
of this 
work, Andr\'e Martin and Paco Yndurain for interesting 
discussions
on the potential model results.
 
\end{document}